# Swinging Crystal Edge of Growing Carbon Nanotubes

Georg Daniel Förster[1,2], Vladimir Pimonov[3], Huy-Nam Tran[3], Saïd Tahir[3], Vincent Jourdain[3,*] and Christophe Bichara[1,*]

[1] Aix-Marseille Univ, CNRS, CINaM, UMR7325, Marseille, France

[2] Interfaces, Confinement, Matériaux et Nanostructures, ICMN, Université d'Orléans, CNRS, Orléans, France.

[3] Laboratoire Charles Coulomb, Univ Montpellier, CNRS, Montpellier, France

*Corresponding authors, e-mails:

 christophe.bichara@cnrs.fr; vincent.jourdain@umontpellier.fr

**Abstract:** Recent direct measurements of the growth kinetics of individual carbon nanotubes revealed abrupt changes in the growth rate of nanotubes maintaining the same crystal structure. These stochastic switches call into question the possibility of chirality selection based on growth kinetics. Here, we show that a similar average ratio between fast and slow rates of around 1.7 is observed largely independent of the catalyst and growth conditions. A simple model, supported by computer simulations, shows that these switches are caused by tilts of the growing nanotube edge between two main orientations, close-armchair or close-zigzag, inducing different growth mechanisms. The rate ratio of around 1.7 then simply results from an averaging of the number of growth sites and edge configurations in each orientation. Beyond providing new insights on nanotube growth based on classical crystal growth theory, these results point to ways to control the dynamics of nanotube edges, a key requirement for stabilizing growth kinetics and producing arrays of long structurally-selected nanotubes.

**Main Text:**

Single Walled Carbon Nanotubes (SWCNTs) are hollow cylindrical crystals which can be seen as rolled up graphene ribbons. They display extreme aspect ratios, lengths up to tens of centimeters and diameters in the nanometer range. Centimeter-long SWCNTs can be grown without defects [1], but a simple pair of pentagonal and heptagonal defects can change their crystal structure [2] which is defined by the so-called Hamada indexes (n,m). Because they display a large number of possible crystal structures which are highly degenerate in energy, the selective synthesis of SWCNTs is specially challenging [3]. In classical crystal growth, the interaction of the growing crystal with its support and the energies of the different facets determine the growth mode and the resulting crystal structure and shape [4]. The synthesis of carbon nanotubes by catalytic chemical vapor deposition (CVD) [5] poses somewhat similar but more complex problems. The catalyst particle is both a support and a reactive interface with the growing tube, and many properties are altered due to the nanometric size of the objects. Thermodynamic and kinetic contributions to SWCNT growth remain relevant at this nanoscale but need to be properly evaluated. For example, the configurational entropy of the tube growing edge in contact with the catalyst, has been shown [6] to drive the stability of the so-called chiral nanotubes, - those that are neither zigzag (n, 0) nor armchair (n, n). A kinetic model of spiral growth predicting a linear dependence of the SWCNT growth rate $G$ with the chiral angle $\theta$ has been proposed previously [7]. This model predicts a smooth and monotonic dependence of the tube length with the synthesis time, partly supported by indirect experimental observations [8,9].

However, our direct *in situ* measurements of the growth kinetics of individual tubes using homodyne polarization microscopy [10] recently evidenced unexpected features contradicting this simplified view: in about half of the cases, sharp jumps between two well defined growth rates are observed, while the tube (n, m) structure remains unaltered. Moreover, we showed that the growth rates measured before and after change ($G_1$ faster, $G_2$ slower) display a proportionality relationship with an average ratio $R = G_1/G_2 \approx 1.7$. Strikingly, measuring individual nanotube growth rates using a different technique, Koyano *et al.* [11] obtained the same 1.7 ratio, before and after exposing growing nanotubes to a water vapor treatment.

In this Article, we present new measurements of individual growth rates for different temperatures (T), precursor partial pressures (P) and catalysts using the optical microscopy method presented in [10]. We show that a similar average ratio between fast and slow rates of around 1.7 is observed regardless of catalyst and growth conditions, suggesting that this is an intrinsic characteristic of SWCNT growth. We propose a simple model based on an enumeration of reactive sites of different categories of tube edges to account for the $R = 1.7$ average ratio of growth rates, and develop it using atomic scale Kinetic Monte Carlo (KMC) simulations whose key parameters are derived from the analysis of the experimental data. Coupled with a phenomenological approach to the bistability of the system, these computer simulations demonstrate that the swings caused by large fluctuations of the nanotube edge in contact with the catalyst are the cause of the striking broken-line kinetic curves observed.

**Results and discussion**

The experiments described in [10] were performed using a Fe catalyst at 1600 Pa partial pressure of ethanol and 1123 K. Here, additional *in situ* data collected at other partial pressures (59, 178, 533 Pa) and temperatures from 1073 K to 1173 K are reported, treated using an automated procedure of video analysis providing both a higher throughput and a better time resolution, as described in Materials and Methods. Figure 1A shows examples of growth rate changes. The time step is 1 s and the length resolution is +/- 0.3 μm. Among the 2350 tubes grown, about one half displayed broken-line kinetics and a total of 1067 rate transitions were observed. These 1067 measured transitions correspond to cases with no detectable change of chirality and no pause. Among them, a subset of 476 data points was obtained at the same ethanol pressure of 533 Pa but at five different temperatures (Figure 1B). Figures 1C and D display the growth rates $G_1$ as a function of $G_2$ measured, respectively, for four different partial pressures at T = 1123 K, and for five different temperatures at P = 533 Pa. The average growth rate ratio R was assessed by two methods: i) by calculating the mean value of GRR distribution, and ii) by linear regression of $G_1$ versus $G_2$ (as shown in Fig. 1C-D). As detailed in SI (Methods and Table S2), depending on the growth conditions, the calculated average proportionality factor varies by 2 % to 20 % around a central value of 1.7 (for the linear fit method) or 2.0 (for the mean ratio method). To confirm the generality of the behavior, we also tested catalysts other than Fe (i.e. Co, Ni, and FeRu): as Fe, these catalysts also displayed growth rate switches. As shown in Figures 1C,D, our data for these catalysts nicely fall along the same trend we measured for Fe, as do the data obtained by Koyano *et al.* also for Fe, but in different conditions [11]. A complete list of growth rates and residence times used in the following is given in Supplementary Data 1. The nature of the substrate (ST-cut quartz) makes it difficult to unambiguously assess the helicity (n, m) of the observed nanotubes by Raman spectroscopy, but the radial breathing modes indicate that the tube diameters are in the range of 1 to 2 nm. The sets of experimental data acquired at five temperatures and P = 533 Pa are rather large, and suggest performing an Arrhenius analysis to evaluate the activation energies $E_1^{act}$ and $E_2^{act}$ associated with the incorporation of carbon atoms at the tube edge in regimes 1 and 2. Because of the large scatter of the data, this process it is not accurate enough to prove they are equal. However, since a solution with very similar activation energies in the two regimes: $E_{act}^1 = 1.13 \pm 0.13$ eV, $E_{act}^2 =$ $1.13 \pm 0.14$ eV is within the uncertainty limits, and for the sake of developing simple models, we assume equal activation energies. In such a case, the observed difference in growth rate would stem from the prefactors, which correspond to entropy contributions, as discussed below.

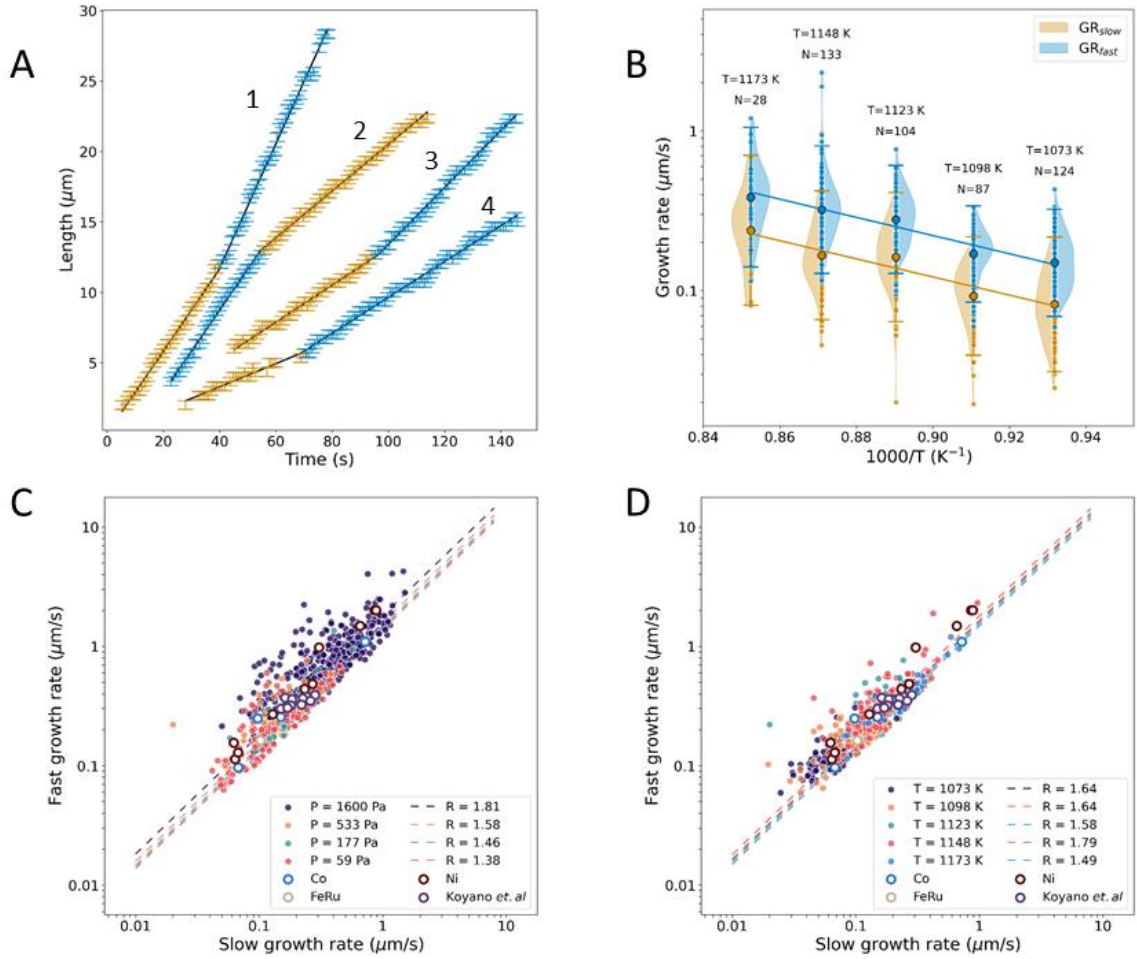

***Figure 1. Experimental growth rate switches of individual SWCNTs from in situ homodyne polarization microscopy.*** *A) Kinetic curves of individual tube growth displaying an abrupt change of growth rate. Error bars are indicated, the time resolution is 1 s and the length resolution is 0.3 µm. Measured growth rates for tubes 1 to 4 are: 0.29/0.44; 0.31/0.17; 0.13/0.20; 0.08/0.13 in µm/s. B) Arrhenius plots of $G_1$ (cyan) and $G_2$ (brown) growth rates. The number of events and the temperatures are indicated in boxes. C) Fast ($G_1$) versus slow ($G_2$) growth rates at each recorded transition, at different partial pressures of ethanol and T = 1123 K. Large open circles correspond to Co, Ni and FeRu and data by Koyano et al.* [11]*. D) Same as in C, but for P = 533 Pa and various temperatures.*

The remarkable stability of $R$ for different catalysts and growth conditions strongly suggests that it is an intrinsic property of SWCNT growth and therefore a key element to understand the bistable growth regime which we evidenced. The very rapid changes in growth rates, on the order of seconds, the lack of correlation between the growth behaviors of neighboring tubes, and the fact that any significant change in the catalyst surface would lead to a different reactivity, while similar activation energies (≈ 1.13 eV) were measured in both growth regimes, rule out changes in the state of the catalyst or catalyst-substrate interactions as the source of the observed dynamic instabilities. The growth model for SWCNTs we build is therefore based on a number of approximations. We start by neglecting the interaction of the tube with the substrate and note that, once a tube is nucleated, it is mechanically stiffer [12] than the Fe nanoparticles under CVD conditions and should impose its structure and dynamics. The catalyst is thus indirectly taken into account through the interface energies between the fluctuating tube edge and the catalyst surface. The observed abrupt kinetic changes show similarities to the reported oscillatory behavior of the interface between a growing nanowire and its catalyst [13,14]. The "jumping catalyst" model [15] explaining the sudden jumps of a catalyst particle from one crystal facet of a growing Si nanowire to another one, observed during *in situ* TEM experiments is a source of inspiration, with the notable difference that, in the following model, it is the edge of the tube that fluctuates and swings between two configurations.

Nanotubes are at least one order of magnitude thinner than nanowires and their edge is a simple line of a few tens of carbon atoms in contact with a particle in the 1-5 nm diameter range. Because of these very small sizes, large fluctuations can be expected. However, observed growth rates remain of the order of $1 \ \mu m.s^{-1}$. For a (10, 10) tube, this would correspond to adding $1.6 \ 10^5$ atoms per second on its 20 edge sites, leading to an efficient averaging in measurements with a time resolution of the order of one second. We previously emphasized the critical role [6] of the configurational entropy of edges of a (n, m) tube in the stabilization of chiral tubes. Edges were considered perpendicular to the tube axis, hence with constant numbers of armchair (2m) and zigzag (n-m) edge atoms.

Here, as depicted in Figure 2 A, B we allow the number of armchair **pairs** $N_a$ and zigzag **atoms** $N_z$ to fluctuate, while still keeping a constant number of edge atoms. The underlying idea is that the different growth rates observed should depend on the orientations of the tube edge with respect to the tube axis, because the number of reactive sites onto which carbon atoms can be attached depends on these edge orientations. This implies that $2N_a + N_z = n + m$ and thus includes the possibility for tube edges to be oblique (tilted) with respect to the tube axis. As a consequence, the number of possible edges increases with the chiral angle and the configurational entropies of the edges of (n, m) tubes are modified. Note that allowing the total number of edge atoms to be larger than (n+m) would lead to a larger energy of the interface which would hence become less stable and display unrealistic edge shapes.

Chirality stability maps calculated with these different assumptions are plotted in Figure S1. Note that within this model (n, 0) tubes cannot grow because nucleating a hexagon on a zigzag edge would create an edge with n+1 atoms, which is not allowed. However, zigzag tubes remain scarce in experimental reports [16, 17, 18], either because they are difficult to identify by photoluminescence [19], the most common screening technique, or because they are truly rare.

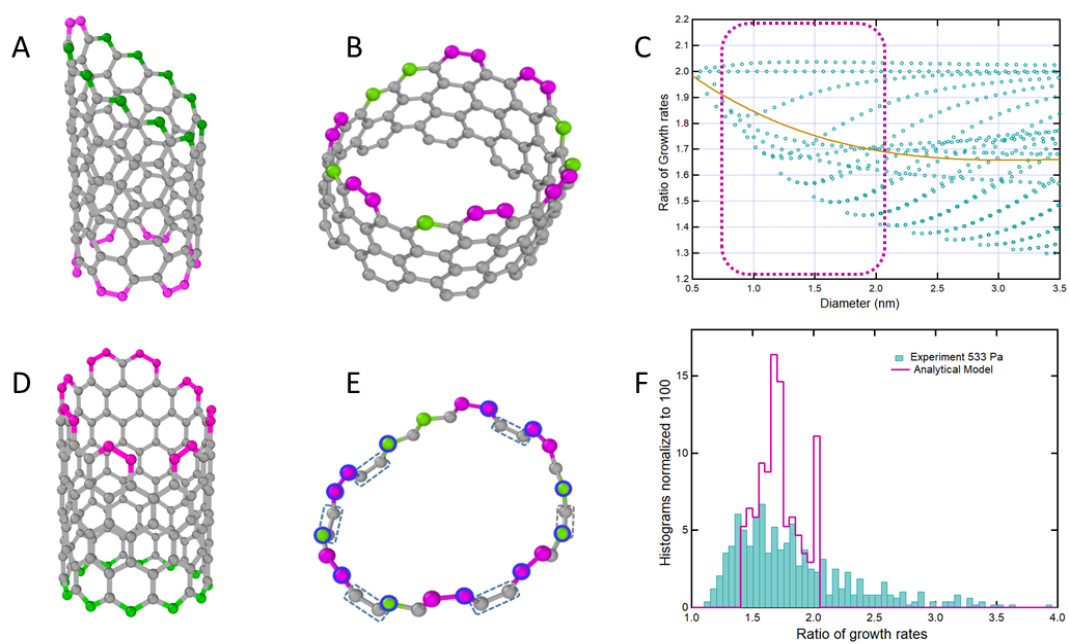

***Figure 2. Basis of the model.*** *A) Armchair tubes, here (6, 6), can have an oblique edge with n+m atoms, here 12, and up to 2n-2=10 zigzag (green) and 2 armchair (purple) atoms. D) Cutting oblique interfaces in zigzag tubes, here (12, 0), leads to more than n+m edge atoms, here 16. B) Example of an edge configuration, showing three-fold coordinated carbon atoms within the tube (grey), zigzag edge atoms (green) and armchair edge atoms (magenta). E) We assume that the only active sites on which incoming carbon atoms can be attached are the edge atoms circled in blue. The number of such active sites is equal to the number of "anti-armchair" atoms in grey boxes, which is in turn equal to the number of armchair edge atoms. The catalyst, not displayed here, sits on top of the tubes, in contact with the colored atoms.* ***Analytical results.*** *C) Ratio of the calculated growth rates $R^{cal} = G_1^{cal}/G_2^{cal}$ in regimes 1 and 2, plotted as a function of the tube diameters. Each dot corresponds to a tube chirality and all possible tubes with diameters below 3.5 nm are included. The brown line is a polynomial fit where all chiralities are equally weighted. F) Histograms of the ratios $R^{exp}$ and $R^{cal}$: experiments at 533 Pa and temperatures from 1073 to 1173 K (cyan) and analytical calculations (magenta).*

We start by developing a simple model to account for the constant ratio of growth rates, by analogy with Schwartz *et al.* [15] and guided by our experimental results suggesting a bistable system with an activated process driving the transitions from one regime to the other. As sketched in Figure 3A, we separate the set of possible edges of a (n, m) tube in two subsets: subset $S_1(n,m)$ includes edges where $N_z < (n-1)$, and subset $S_2(n,m)$ where $N_z \geq (n-1)$. We note that $2 \leq 2N_a \leq 2m$ and $(n+m-2) \geq N_z \geq (n-m)$. The limiting value $(n-1)$ separating the two growth regimes is thus simply the average between $(n+m-2)$ and $(n-m)$. In the case of graphene, growing edges could be observed and the relative stability of armchair and zigzag edges has been evaluated, with contrasted results [20, 21]. Here, since no direct observation of the tube/catalyst interface during growth is practically possible, the validity of the model will be assessed by its ability to reproduce available experimental data.

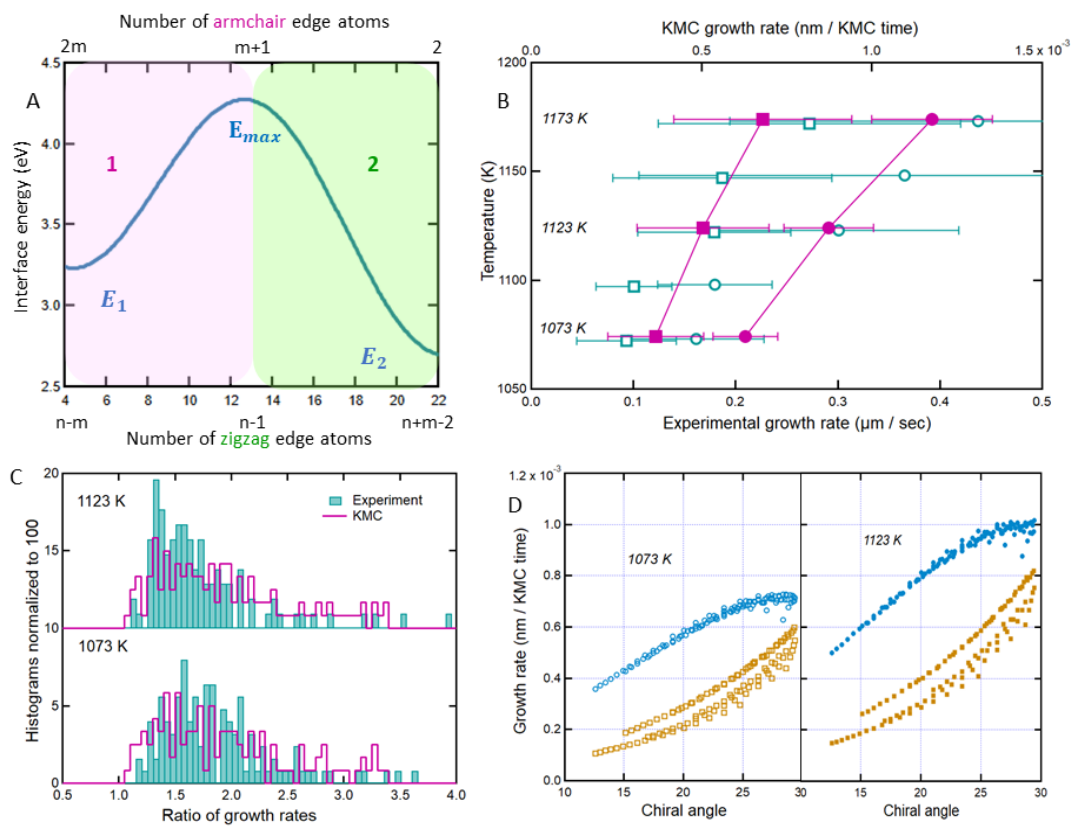

**Figure 3.** *A) Scheme of the bistable energy model*: *Regime 1 corresponds to edges where $N_z < n - 1$, and regime 2 corresponds to $N_z \geq n - 1$, the splitting value being $n_z = n - 1$. Interface energies depend on (n, m). Orders of magnitude of $E_1$, $E_2$ and $E_{max}$ can be obtained from the experiments. The curve is a plot of the interface energy (Eq. 6) including the bistability term (Eq. 7), for a tube (14, 10) whose energy parameters are presented in the main text.*

*Results of KMC simulations restricted to each edge set. B) Growth rates in regimes 1 (circles) and 2 (squares). Cyan, open symbols are growth rates measured at P=533 Pa and 5 temperatures, with experimental standard deviations. Magenta, full symbols are calculated growth rates, error bars are the spread of the data for 120 tube chiralities between 0.8 and 2.0 nm diameter. Full lines are a guide to the eyes. C) Ratio of the calculated growth rates compared to experimental ones at 533 Pa and 1073 K (bottom) and 1123 K (top) panels. D) Fast (regime 1, blue) and slow (regime 2, brown) growth rates calculated at 1073 and 1173 K, plotted as a function of the chiral angles. Interestingly, the positive curvature indicates a more pronounced kinetic selectivity towards large chiral angles in regime 2 than in regime 1.*

The present model derives from the KMC simulations we developed recently [22]. The simplest way to grow a defect-free tube is to add 2 carbon atoms at locations, "cosy corner" [7] or facing "anti-armchair" sites, where they complete the 4 edge sites to form a hexagon, as shown in Figure 2 B, E and in reference [22]. One readily sees that the number of such active sites is equal to the number of armchair sites. The number of edge configurations with i armchair pairs $A(n, m, i)$ and the total number of edge configurations accessible to a (n,m) tube $N(n, m)$ are:

$$A(n, m, i) = \frac{(N_a^i + N_z^i)!}{N_a^i! \; N_z^i!} = \frac{(n + m - i)!}{i! \, (n + m - 2i)!} \qquad N(n, m) = \sum_{i=1}^{m} A(n, m, i) \qquad Eq. 1$$

This number of edge configurations and the fact that the edge energy depends on the number of armchair and zigzag edge atoms lead to a different configurational entropy as compared to our previous model [6]. Figure S1 shows that the entropy increases for chiral angles up to ~18° and then stays almost constant, leading to larger stability domains for tubes on the armchair side. Adding two carbon atoms at these favorable locations is a complex process that depends on the nature of the catalyst and CVD conditions, as well as the local atomic-scale environment. The problem of how edges grow step by step by additions of C dimers has been tackled by Dümlich *et al.* [23]. We assume that each addition can be cast into a Transition State Theory (TST) framework, a reasonable assumption used in [22]. In such a case the escape rate, defined as the rate of successful addition of a C$_2$ dimer on a site $\lambda$, per time unit is:

$$k_\lambda = v_0 \, exp(-(\delta E_\lambda - \delta \mu_C \delta N)/k_B T) \qquad Eq. 2$$

In this expression $v_0$ is the number of attempts to jump over the barriers $\delta E_\lambda$ per unit of time, setting the time scale, $\delta \mu_C$ is the carbon chemical potential, driving force for the growth, and $\delta N = 2$ for the addition or removal of two carbon atoms. One needs to sum over all available escape rates $k_\lambda$ to obtain the time increment at each kinetic step. For the sake of developing a simple model, and in agreement with the equal values of activation energies $E_{act}^1 = E_{act}^2 = 1.13$ eV measured in both growth regimes, we make the additional assumption that all energy barriers to the addition of a pair of C atoms ($\delta E_\lambda$) are equal, whatever the zigzag or armchair nature of addition sites. The total rate for a given edge with 2i armchair sites is then equal to the number of armchair sites multiplied by a constant rate term: $k_{tot} = 2i \, k$. The corresponding time increment for each addition of a carbon dimer is:

$$\delta t_\alpha = k_{tot}^{-1} = (2i \, k)^{-1} = \frac{exp\left((\delta E_\lambda - \delta \mu_C \delta N)\big/ k_B T\right)}{2i \, v_0} \qquad Eq. 3$$

Averaging over the number active sites 2i of the various edge configurations leads to an average time increment $\delta t_\alpha(n,m)$ for regimes $\alpha = 1$ and $\alpha = 2$, where the sums run over the relevant edge subsets $S_\alpha(n,m)$:

$$\delta t_\alpha(n,m) = k^{-1} \frac{\sum_{S_\alpha(n,m)}(2i)^{-1}A(n,m,i)}{\sum_{S_\alpha(n,m)}A(n,m,i)} \qquad Eq.\,4$$

This leads to two sets of data depending on the parity of m, as shown in Figure S2. The calculated growth rates are $G_\alpha^{cal} \sim 2\delta t_\alpha^{-1}$, $\alpha = 1$ or $2$ and their ratio is:

$$R^{cal} = \frac{G_1^{cal}}{G_2^{cal}} = \frac{\delta t_2}{\delta t_1} \qquad Eq.\,5$$

As we assume that all barriers have equal heights, the thermodynamic factor k cancels out and we thus demonstrate that the ratio $R^{cal}$ depends only on the numbers of carbon integration sites within the accessible subset of edges in each regime. Figure 2 C displays the calculated $R^{cal}$ for all chiralities with diameters below 3.5 nm, and Figure 2 F shows a comparison with experiments performed at P=533 Pa. Average experimental values of growth rates and residence times in each regime are available in Supplementary Table 1. Figure S2A shows that the peak around 2 results from tubes with chiralities (n, 2) or (n, 4). Since such a sharp feature is absent in the experimental signal, we assume that these near-zigzag chiralities are rare in our growth conditions, as was already frequently reported by other groups. In the following, KMC calculations and comparison with experiments are thus done for a set of 120 tubes with m ≥ 4 and diameters below 2.0 nm. Despite its extreme simplicity the model correctly yields $1.4 \leq R^{cal} \leq 2.0$ with averages around 1.7 for tubes in the 0.75 - 2.0 nm range. It should be noted that the average assumes an equal weight for all chiralities in this range and thus does not take into account possible unequal abundances of chiralities in the experimental samples. It also explains why the experimental growth rate ratios are essentially independent of the catalyst and growth conditions [11, 10]. The tail of the experimental growth rate ratio distribution for values greater than 2 is not properly represented in our very simple analytical model. This is a reason why, in the following, we develop a more accurate kinetic Monte Carlo modeling that corrects for this feature and enables a more detailed investigation of growth kinetics.

The next step of our analysis is a phenomenological approach to reproduce broken line kinetics within the assumptions presented above, which implies moving from analytical to atomistic modeling, using Kinetic Monte Carlo (KMC) simulations [22]. These simulations are based on a description of interatomic interaction energies that can be adjusted to match experimental results and take temperature into account, allowing for more accurate comparisons. The main features of the atomistic energy model are the following. We use a simple lattice gas model for each tube chirality. The carbon chemical potential $\mu_C$ acting as an external field, the lattice is gradually filled with C atoms using a semi grand canonical algorithm [22, 24] and we compare the edge stabilities and growth kinetics obtained for different chiralities. Edge atoms are either zigzag or armchair, with energies $E_a$ or $E_z$ respectively. The energy of the 3-fold coordinated atoms in the body of the tube is set to zero. With this energy reference, the total energy is reduced to the interface energy which writes:

$$E_{int}(n,m,i) = 2N_a E_a + N_z E_z + E_{curv} + E_b \qquad Eq.\,6$$

The curvature energy term $E_{curv}$ is used only in the thermodynamic modeling [6]: since simulations are done on fixed lattices, it is constant for each run and is thus neglected because only energy differences between configurations matter in the KMC process. $E_b$ is a correction term that favors a bistability by taking into account the possible mismatch between the tube edge and the catalyst surface. As explained below, it is used only to simulate the transition from one regime to the other, in the second step of our KMC simulations. Energy barriers $\delta E$ are activation barriers for the addition of a pair of C atoms at the reaction sites, to complete a hexagon. As described in [22], they could be fine-tuned by using additional energy terms taking into account the zigzag or armchair nature of the C addition sites. However, for the sake of simplicity, they are considered constant in the following simulations.

To calibrate the interface energy of armchair and zigzag atoms ($E_a$, $E_z$) of our KMC simulations, it might seem appealing to use DFT calculations of the formation energies of tube/catalyst interfaces, as performed for Fe and other catalysts by Ding *et al.* [25] and Hedman [26]. However, extrapolating such data calculated for simple interfaces to the large number of possible interfaces, obtained by mixing armchair and zigzag edge atoms keeping the total number of edge atoms constant, might be misleading. We retain that interfaces of armchair tubes with Fe, Co, Ni are less stable than zigzag ones, and thus assume that the average interface energy in regimes 2 is lower than in regime 1 ($E_2 \leq E_1$), as sketched in Figure 3A. This is qualitatively confirmed in Supplementary Table 1 which displays the average duration of growth segments, or residence times, $t_1^r$ and $t_2^r$ in regimes 1 and 2. We observe that the latter is systematically longer, with ratios $1.19 < \frac{t_2^r}{t_1^r} < 1.67$ implying that the barrier $E_{max}$ seen from 2 is higher than seen from 1. Because of the limited number of possible edges, no bistability is theoretically possible for $m = 0$ or $m = 1$, and practically for $m < 4$. Separating the possible edges in two subsets in the simplified analytical model was essential to obtain two different growth regimes and reach an agreement with experimental data. In the following, we use two different approaches to investigate the origin and characteristic features of the growth rate changes.

As a direct development of the analytical model, we start by using KMC simulations sampling separately the edge subsets $S_1(n,m)$ and $S_2(n,m)$. In this case, growth is performed in either regime, through an appropriate selection of attempted insertion / removal moves: any KMC move that would carry the system over to the other regime by passing the threshold of the number of permitted zigzag atoms is rejected. Figure S3 displays the distribution of the probabilities of the different edges of tubes belonging to the $(n + m) = 18$ family, obtained from standard and biased KMC simulations. The above mentioned DFT calculations and experimental data of $G_1$, $G_2$ and $R$ at P=533 K and T=1123 K set orders of magnitude for the parameters of the KMC simulations. Some parameters such as $\mu_C$, $E_a$, $E_z$ are not directly accessible from these experiments, some others, such as $\gamma$, $\Delta 1$ and $\Delta 2$, in principle depend on the tube chiralities. Since we do not know the chiral distribution of the experimental samples, a direct fit of these parameters is not possible. We allow some flexibility to obtain a set values compatible with each other in a reasonable range and took $E_a = 0.14$ eV and $E_z = 0.11$ eV, as well as $\delta E = 0.9$ eV. The carbon chemical potential was set to $\mu_C = 0.10$ eV/at. in all KMC simulations. Figures 3 B, C show that a satisfactory agreement is achieved between experimental and calculated growth rates $G_\alpha$ and their ratio $R$ at all temperatures. Error bars on calculated values correspond to the spread of the results for all tubes between 0.75 and 2.0 nm, while experimental ones correspond to fluctuations obtained over the unknown experimental chirality distribution. Note that the model predicts that the average growth rate ratio depends on the actual nanotube population and interface energy values: the 1.7 value therefore slightly varies with the growth conditions. The data in Figure 3 B are presented in an extended, more legible form in Figure S4. The comparison of the KMC time and the real time ($1.5\ 10^{-3}\ nm \cdot t_{KMC}^{-1} \approx 0.5\ \mu m\ s^{-1}$) on the x-scales enables us to estimate the KMC time unit at about $3.0\ 10^{-6}$ seconds of real time. Figure 3 D displays the growth rates $G_1^{KMC}$ and $G_2^{KMC}$ calculated by KMC in each regime plotted as a function of the chiral angle. This calculated dependence of growth rates on SWCNT chirality is an important prediction of the model whose experimental validation requires the determination of the chirality of a statistically large number of SWCNT with different chiralities, which is currently not included here, due to the difficulty of accurate chirality assessment of individual SWCNTs on ST-cut quartz substrates by Raman spectroscopy and optical methods in general.

Another important aspect, evidenced in Figures S5 A, C, is that the growth mechanisms in regimes 1 and 2 are qualitatively different. Considering the position of the edge atoms along the tube axis as their height and remembering that the catalyst nanoparticle sits on top of this interface, we can see that in regime 1 edge atoms are mostly armchair (pink) and fluctuate around an average height that keeps moving upwards during growth. Since the number of growth sites is equal to the number of armchair pairs $N_a$, it is by definition larger in regime 1, which grows faster. The larger number of possible edge structures contributes to a larger configurational entropy and a rough interface is formed. In regime 2, edge atoms are mostly zigzag (green) and incoming dimers must climb from a low point, randomly choosing a clockwise or anti-clockwise direction. Depending on the interface energies $E_a, E_z$ and energy barriers $\delta E$, renucleating a growing step from the low point in regime 2, after the previous one merged with the topmost armchair site, can take time, therefore making regime 2 even slower. One can even imagine situations where growth would be temporarily or permanently blocked if this renucleation is slower than the flux of incoming carbon atoms, which could then either diffuse inside the catalyst and modify the feedstock decomposition or encapsulate the nanoparticle.

In the above discussions, we introduced the bistability that is observed experimentally, by rigidly dividing the edge configurations into two groups. Under this rather strict assumption, the analytical model and KMC simulations restricted to either group were used to reproduce and explain the main experimental features. In order to show that the sharp changes of growth rate are caused by sharp changes of the edge structure we now proceed to directly simulate the transition from one regime to the other, by introducing a bistability into regular KMC simulations, which means sampling all possible edges within a run. This could be done by taking into account the tendency for armchair and zigzag edge atoms to separate from each other, as already established by Bets *et al.* [27] using first principles calculations of tube / catalyst interfaces in which the numbers or armchair and zigzag edge atoms is constant, but their ordering varies. However, the additional flexibility offered by variable numbers of armchair and zigzag edges, suggests that the easiest way to separate them is to create an oblique (tilted) edge, when possible. A possible solution is then to include a smooth correction term to the interfacial energy (Eq. 6) in the following form:

$$E_b = \gamma m [\sin(\pi(N_z - n + m))/2(m-1)]^2 \qquad Eq.\,7$$

$E_b$ is a squared sine function, equal to 0 for $N_z = n - m$ and $N_z = n + m$-2 with a maximum $\gamma m$ for $N_z = n - 1$. This empirical term, which induces a bistability and a tendency to separate armchair and zigzag atoms is a way to compensate for the oversimplified nature of the interface energy model, which neglects the atomic roughness of the catalyst surface, its geometry and mechanical properties. As shown in Figure S6, including only segments followed by an actual growth rate jump, we can use the $t_\alpha^r$ data at different temperatures to obtain $\Delta_1 = E_{max} - E_1 = 1.11 \pm 0.19$ eV and $\Delta_2 = E_{max} - E_2 = 1.22 \pm 0.11$ eV. They are fitted against 133 (regime 1) and 343 (regime 2) experimental lifetime data points at 5 temperatures taken out of our 2 x 476 points global data set, leading to a limited accuracy. Note that these values of $\Delta_1$ and $\Delta_2$ depend on the chirality distribution through equations 6 and 7. Since we don't know the actual experimental distribution, we use these lifetime data to set an order of magnitude for the parameter γ. The value of γ sets the amplitude of $E_b$ which, together with the interface energies, temperature and chemical potential, controls the number and amplitude of growth rate jumps. We found that taking $\gamma = 0.14$ eV led to a good qualitative agreement with the experimental data. KMC runs with $2.10^6$ KMC steps were performed for tube diameters below 2.0 nm, resulting in tube elongations up to 20 $\mu m$, close to experimental values. Figure 4 A displays the fluctuating fractions of zigzag atoms for tubes belonging to the (n, n-4) to (n, n) families, obtained with the same parameters as the previous KMC results and $\gamma = 0.14$ eV. For a given set of KMC parameters, not all tubes display a broken-line growth kinetics, as this depends on the values of the interface energies and KMC growth parameters. Figure 4 B displays the tube length and the fraction of zigzag atoms as a function of the growth time. It demonstrates that edges with large fractions of zigzag atoms correspond to slower growth (regime 2), while those with fewer zigzag edge atoms lead to faster growth (regime 1), as postulated previously. The gradual washing out of the kinks with increasing temperature, evidenced in Figure 4 C, is expected for a thermally activated process. Here and in our previous experiments [10] the observed number of kinks is variable. In our bistable system modeling, this number depends on the value of the correction term $E_b$ and hence $\gamma$. The results shown in Figures 4 A, B, C correspond to a trade-off between the number of kinks observed, the distribution of the growth rate ratios and the general agreement with the experiments. To demonstrate the correlation between the growth rate and the edge orientation changes we display here growth sequences with many kinks. Figure S5 B displays a series of configurations showing that the transition between the two regimes takes place when at least two armchair pairs remain separated from each other on the edge, in a transient state where the edge adopts a saddle shape. Another growth sequence of a (13, 9) tube is presented in Figure S7 A, - different from the one presented in Figure 4B because another sequence of random numbers was used -, for which we analyze the flatness of the tube edge. During growth the interface fluctuates, and is flatter in regime 1 than in regime 2. Figure S7 B shows that larger deviations from a flat interface are observed in the transition region, associated with the saddle shape of the interface visualized in Figure S5 B. Such deviations from flatness were observed in KMC simulations using either $E_b$ (Eq. 7) or including an ordering energy term favoring the separation of armchair and zigzag edge atoms at the interface [27]. For catalysts such as Fe, whose carbon solubility is important, the edge of the tube is attached to a metal nanoparticle generally larger than the diameter of the tube [28]. In such a case the part of the catalyst in contact with the tube can be considered as locally flat. A saddle shaped interface at the transition is then less stable than in either steady state, because either some of the carbon-metal bonds are weakened, in the case of a rigid, nearly flat catalyst, or the catalyst is deformed if these bonds remain strong enough to pull the metal surface and deform it, as discussed by Qiu *et al.* [29].

Using the standard deviation of the difference between the fractions of armchair and zigzag edge atoms ($N_z - 2N_a$) as an indicator of the number and amplitude of kinks, Figure 4 D shows that broken-line kinetics are observed in a limited domain of close-to-armchair chiralities, with the present interface energy parameters. It is important to note that this domain of growth instabilities shifts if the interface energies are different. For most catalysts [26], $E_a - E_z > 0$. Figure S8 shows that the instability domain tends to shift towards central chiral angles, when this difference becomes larger. On the contrary, if the armchair interface energy is more stable, our model predicts that instabilities should disappear.

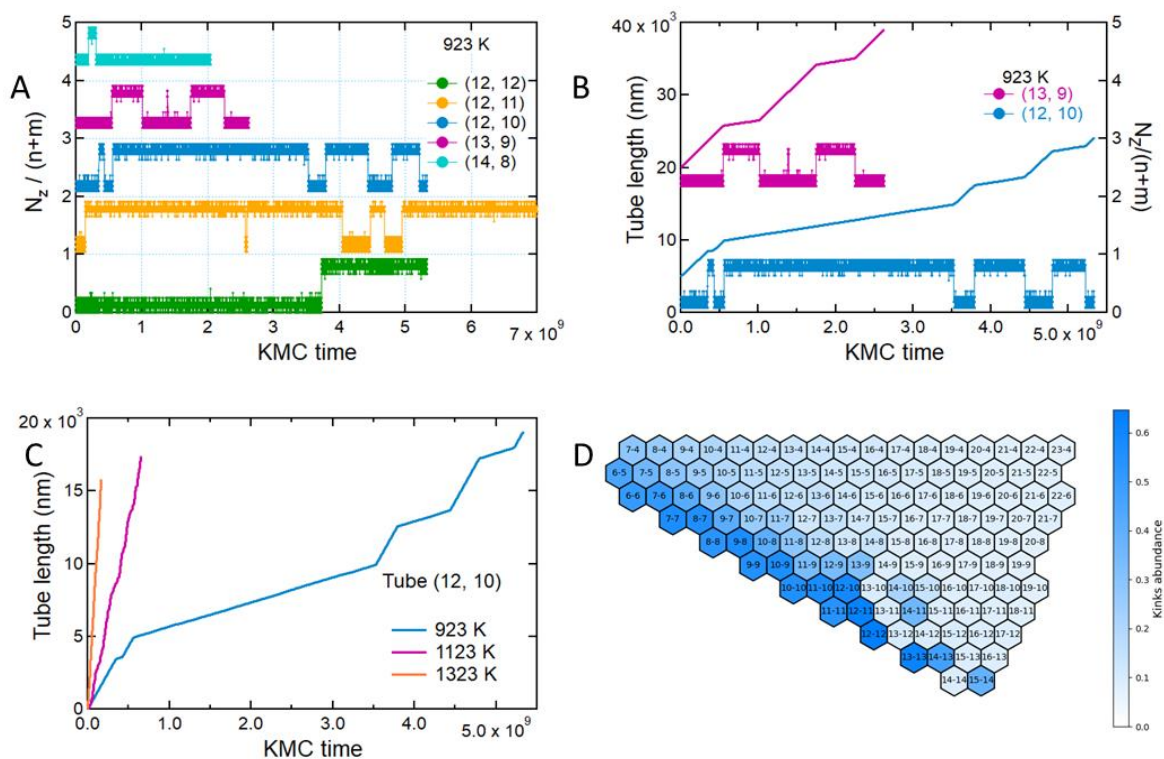

***Figure 4. KMC modeling of growth instabilities***. *A) Fluctuating fractions of zigzag atoms recorded for (12, 12), (12, 11), (12, 10), (13, 9) and (14, 8) tubes at 923 K. 25000 points were recorded for each chirality; B) Sharp growth rate changes ("kinks") are visible for (13, 9) and (12, 10) tubes, associated with growth rate changes. Slower growth rates correspond to large fractions of zigzag edge atoms $N_Z/(n + m)$, regime 2, faster ones to large fractions of armchair atoms, regime 1; C) As expected for a thermally activated processes, kinks tend to wash out with increasing temperatures; D) The standard deviation of the difference between the fractions of armchair and zigzag edge atoms ($N_Z - 2N_A$), plotted as a function of (n, m), is a good indicator of the number and amplitude of kinks. In these conditions, kinks are localized close to the armchair edge.*

**Conclusions**

The present modeling and KMC simulations demonstrate that fluctuations of the tube/catalyst interface between different orientations with respect to the tube axis, leading to different growth mechanisms, is responsible for the sharp growth rate changes evidenced by *in situ* measurements of the growth kinetics of individual tubes. Such an uncommon behavior derives from the ability of carbon nanotubes to form two types of edge atoms with different energies, from their very small diameter, and from interface energies in an appropriate range to be balanced by the edge configurational entropy. It is thus typical of the growth of SWCNTs. Yet, a comparison with classical surface science and crystal growth [4] highlights clear similarities. SWCNTs are crystals whose interface with the substrate/catalyst is a line of atoms, instead of a surface in a usual crystal. The edges cut with different angles with respect to the tube axis can be seen as equivalent to vicinal surfaces. The moving carbon incorporation sites in regime 2 are similar to step flows observed during the growth of a GaAs nanowire [30]. The transition between regimes 2 and 1, displays some analogies with a roughening transition, gradual because of the very small size of the interface, and driven by the larger entropy of the interface in regime 1. This theoretical approach shows that both stability [6], which matters for nucleation, and growth kinetics are driven by the dynamical disorder at the tube-catalyst interface.

From a practical point of view, abrupt and stochastic changes in growth rates, concerning a specific but unknown range of chiralities, are obviously detrimental to a controlled and selective growth of SWCNTs. For instance, the recent work by Liu *et al.* [31] underlines the need for a better control of growth kinetics to trigger timely the electro-renucleation [32] which switches growing SWCNTs to semiconductors. The understanding that emerges from our current data analysis and modeling can help circumvent this problem. A possibility is to strongly favor one of the growth regimes by making the transition more unlikely. Looking at Figure 3 D, growing tubes in regime 2 might be more selective, though slower. This could be achieved by lowering the temperature which results in longer residence times in regime 2, see Table S1, and/or using mechanically rigid, and thus non-liquid, catalysts. The results of Yang *et al.* using $Co_7W_6$ catalyst [33] or the use of high melting point elemental catalysts [34, 35] can be interpreted as following this direction. The influence of experimental parameters (such as the water concentration but also the gas phase composition in general, the substrate, the nature, size, shape and physical state of the catalyst particle) on the stability and transition probability between these two edge states is an important point which requires further experimental work and modeling. With this improved understanding, further strategies for an improved selectivity should be developed to stabilize the growth kinetics.

# Supporting Information

## Materials and Methods

Experimental details are similar to those described in our previous work *(10)*. We remind the key points.

***Catalyst preparation.*** Growth substrates were ST-cut quartz samples. Since quartz is birefringent, a wedge with optical quality was created to avoid optical reflection with uncontrolled polarization from the back side. The quartz substrates were annealed at 900°C for 8 h. Etched optical marks were created by optical lithography followed by plasma etching. A photoresist Shipley Microposit S1818 doped with a solution of catalyst salt (10 mM) in methanol was used to create catalyst patterns by optical lithography followed by 5 min of calcination in air at 700°C.

***CNT growth.*** CNT growths were performed in a home-made CVD setup consisting of a gas/vapor supply system connected to an *in situ* optical cell (Linkam TS 1500). The cell contains a ceramic sample cup which is used to heat the sample. A first argon line was injected in a bubbler with ethanol at 0°C and a second argon line was used to control the partial pressure of ethanol between 8 and 1600 Pa. The total gas flow in the cell was 1422 sccm. Growth temperatures were 850-950°C. $O_2$ and $H_2O$ concentrations at the CVD outlet were constantly monitored using dedicated inline sensors. These conditions led to the lattice-oriented growth of long individual CNTs (typically 3-80 μm in length) extremely parallel to each other due to the strong alignment effect of the monocrystal surface.

***In situ optical microscopy***. The optical setup is detailed in references (10, 34). Shortly, a supercontinuum light source (Fianium SC-400-4, 2 ps pulses, 40 MHz) is used to illuminate the sample. The diameter of the incident beam is controlled by a lens-based beam reducer to improve the polarization conservation of the microscope objective by reducing the numerical aperture. To improve the uniformity of the beam, a 50-μm pinhole is used as spatial filter. The beam is then polarized by a first Glan-Thompson polarizer (P1) along the axis x. A 80/20 beam-splitter allows illuminating the sample and collecting the signal. A long-distance microscope objective (Nikon Plan Fluor ELWD 20x 0.45 C L) chosen for its polarization conservation is used to illuminate the sample. A second Glan polarizer (P2) along the axis y analyses the light from the sample which is then collected by a camera (Hamamatsu digital camera c11440 ORCA-Flash4.0 LT). During syntheses, a shortpass filter with cutoff at 700 nm was positioned between the analyzer and the camera to suppress the black-body radiation emitted by the crucible. Because homodyne detection relies on the modulation of a strong local oscillator signal, the exposure time was set to around 25-42 ms to avoid saturating the camera and the frames were then averaged over a time step of 1 s. This represented the optimal tradeoff for imaging CNT growth with μm-resolution at typical growth rates and very good signal/noise ratio.

***Automated video analysis.*** The extraction of the CNTs growth kinetic data was performed using AI assisted system in three steps. First, CNTs were located at every frame of the video using neural network with Mask-RCNN architecture trained for this task. Next, segments from each frame were attributed to each other to track the nanotubes movements. Then the information about locations of the nanotubes were extracted and fitted using linear fits to obtain its growth rates.

***Identification of chirality changes.*** As previously reported (see reference 10 of the main text), the vast majority of chirality changes are visible in *in situ* videos because they cause a neat and abrupt change of optical contrast: all such cases were removed from the experimental database used in this work. The Raman analyses we performed show that, after this selection, only a small percentage of chirality changes remain (~14 %). As also reported in reference 10, chirality change cases display a behavior quite similar to the cases of growth rate changes with no chirality change (i.e. displaying peaks at GRR > 1.3). As shown in Figure S9, the experimental distribution of GRR could not be satisfyingly reproduced by a model considering that the growth rate is simply proportional to the chiral angle. The best agreement with the experimental data is achieved with the model allowing for growth regime switches and chirality changes with chiralities within a small diameter range. This suggested that the actual edge configuration may also influence the growth rate, in addition to the influence of the nanotube chirality. The few remaining chirality change cases therefore have a negligible effect on the distributions, let alone the mean growth rate evaluation of around 1.7.

***Calculation of the proportionality factor.*** Two methods were used to assess the average proportionality factor between experimental G1 and G2 values. The first one was to calculate the growth rate ratio (GRR) G1/G2 of each event and to calculate the mean value of this distribution of GRR values. The width of these distributions (e.g. median absolute deviation) is in the range of 0.3. The second method (illustrated in Fig. 1C-D) was to plot G1 versus G2 values and to perform a linear regression (y=ax) to assess the slope a: the uncertainty on the fit value can be estimated from the one standard error, which corresponds to the range the fitted value is estimated to be with a confidence level of 67 %. Note that this one-standard error on the fit output does not represent the width of the GRR distribution.

Both methods were applied to each dataset (specific (T,P) condition, all T at constant P, all P at constant T, all T and P) and the results are summarized in Table S2. It can be seen that, depending on the growth conditions, the calculated average proportionality factor varies by 2 % to 20 % around a central value of 1.7 (for the linear fit method) or 2.0 (for the distribution mean method). Note that, despite the rather broad distributions of values (MAD ~ 0.3), the standard error on the mean or the slope (which scales as the square root of N) is small (typically a few %) due to the large size of the samples.

**Figure S1**

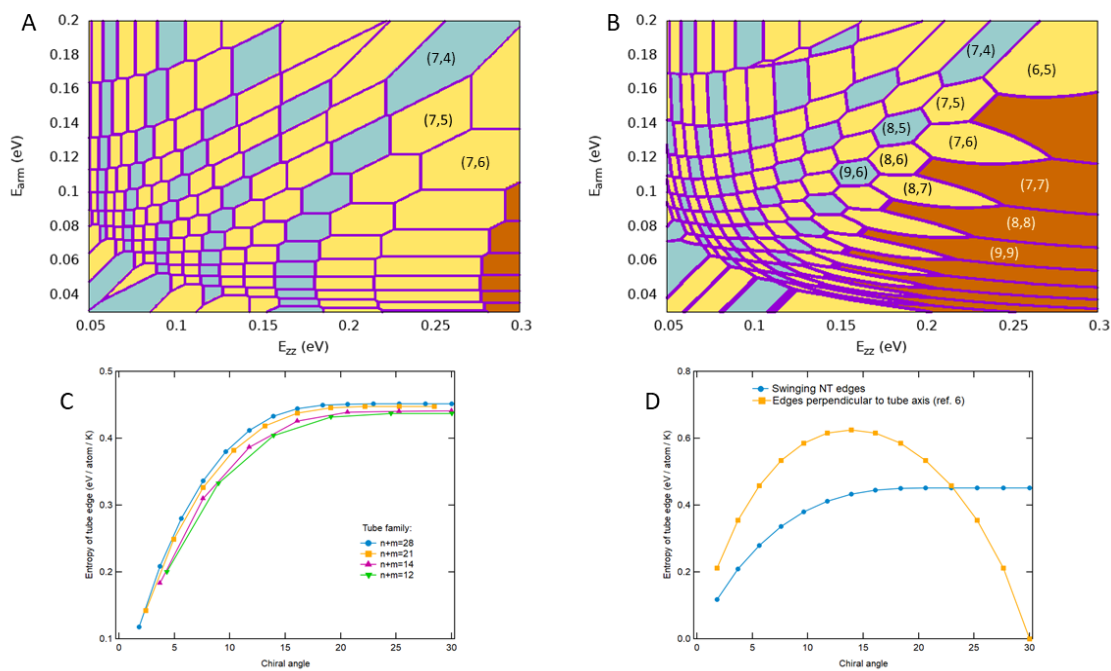

Chiral stability maps calculated at 1123 K assuming different ways to cut the tube interface in contact with the catalyst, hence two different configurational entropies calculated using different partition functions, microcanonical in case A, canonical in case B. A) Tubes are cut perpendicular to their axis, as done in *(6)*. B) Entropies and free energies calculated allowing oblique cuts of the edges, enumerated in Equation 1 of the main text. The entropy is larger on the armchair side in this case, leading to broader stability domains for tubes with large chiral angles. The main message *(6)* that chiral tubes are stabilized by the configurational entropy of the edge still holds. C) Configurational entropies calculated within the current odel for different constant n+m tubes families. D) Comparison of configurational entropies of edges of the n+m=28 tubes family, in the current and previous *(6)* models.

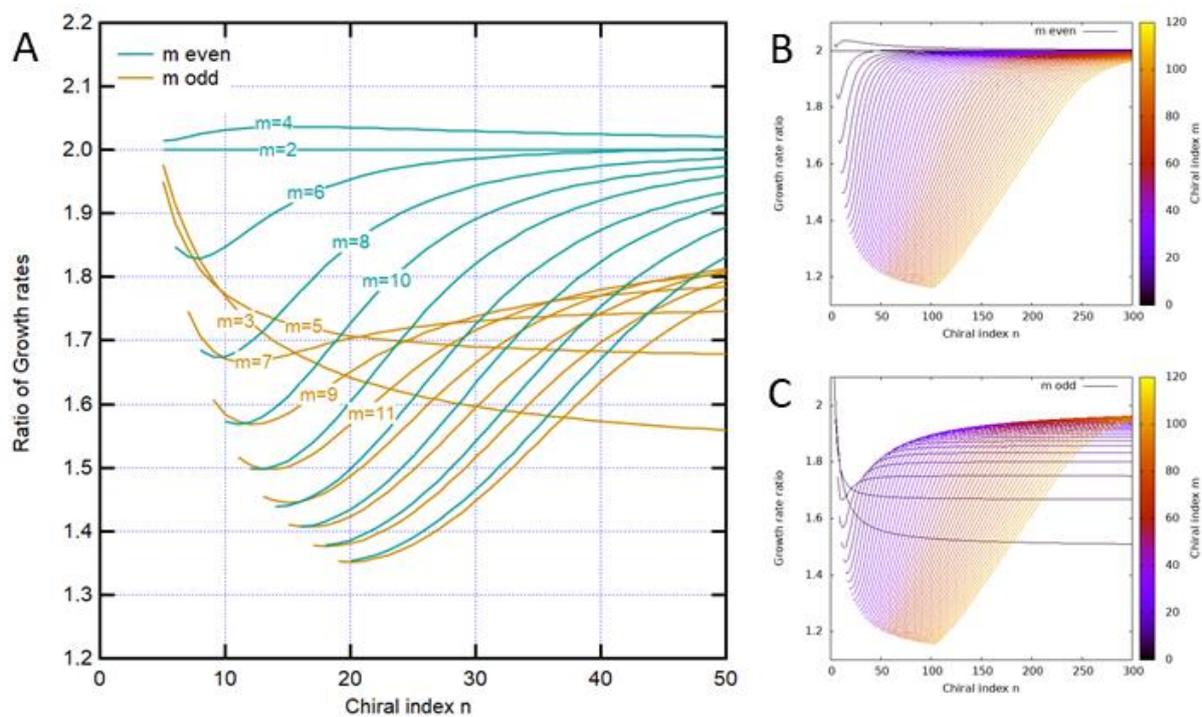

A) Detailed results of the analytical calculation of the ratio of growth rates using Equation 5 in main text. Values of $R$ for tubes (n, 2) and (n, 4), which are expected to appear as a small fraction in the experimental chiral distributions, are centered around 2. Ratios of growth rates calculated for large values of (n, m) for even (B) and odd (C) values.

**Figure S3**

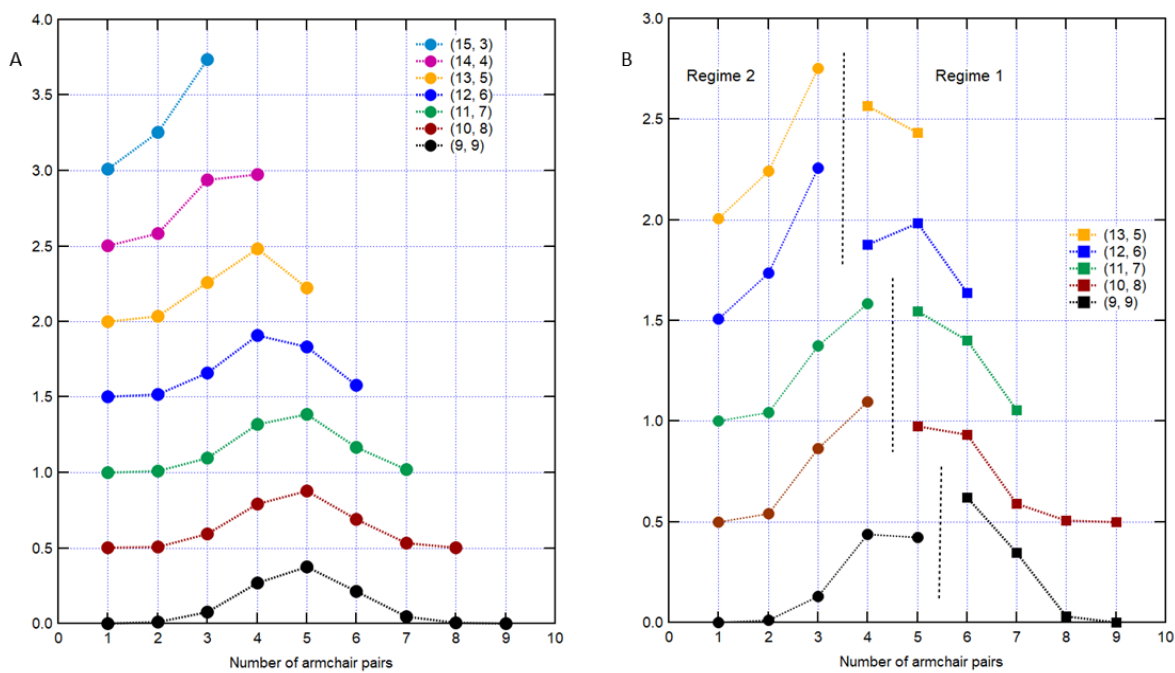

Distributions of the probabilities of different edges with $i = 1, m$ armchair pairs obtained from KMC simulations with different conditions for tubes belonging to the $(n + m) = 18$ family. All distributions are y-shifted for readability. A) KMC simulations sampling the full range of edge configurations. B) Biased KMC simulations, corresponding to Figures 3 B, C, D, where the distributions in each regime are normalized to 1.

**Figure S4**

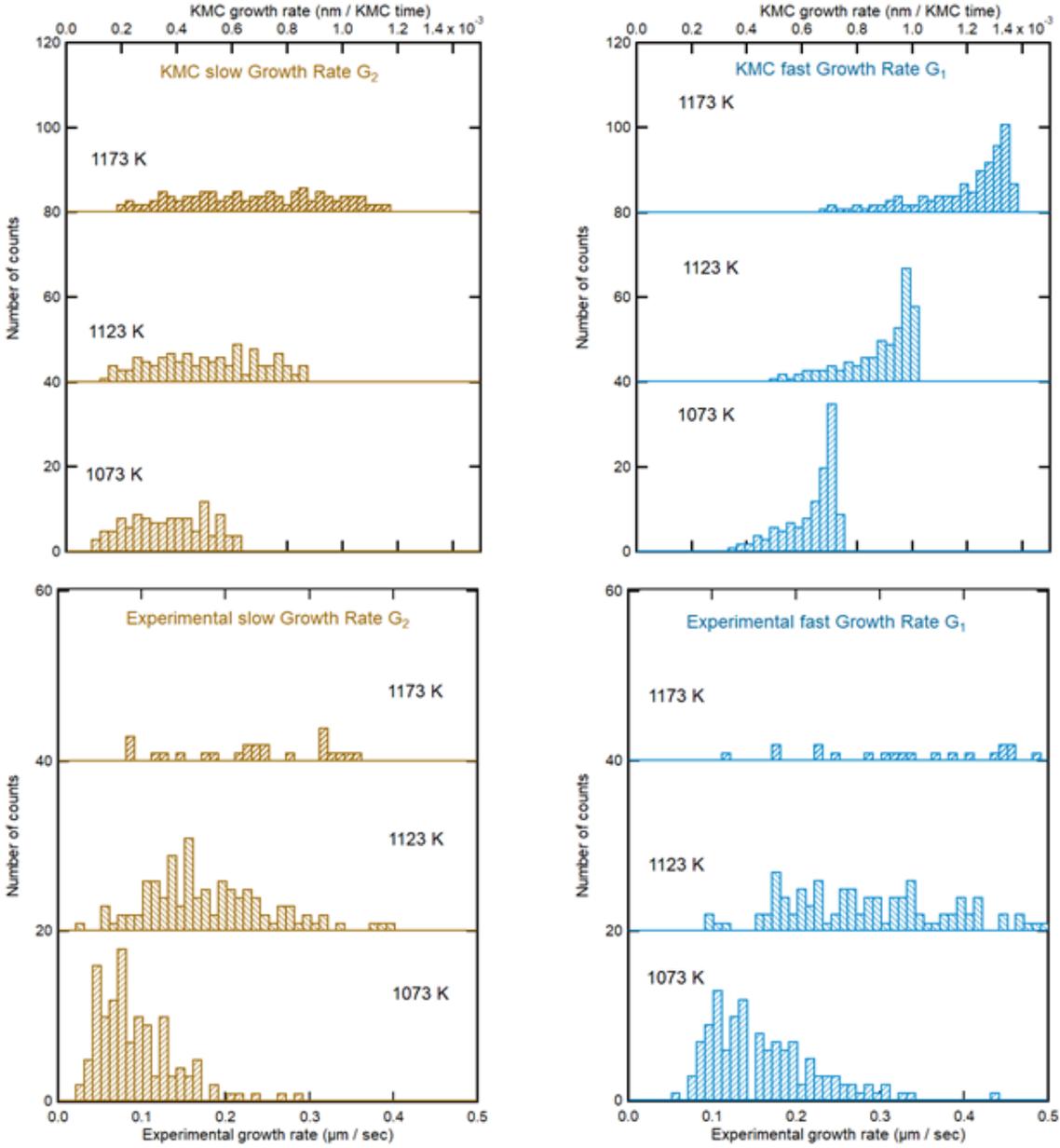

Detailed presentation of fast (G1) and slow (G2) growth rates calculated by our KMC simulations (top row) and experimentally measured (bottom row). These data are the same as those in Fig. 3B, repeated for better legibility

**Figure S5**

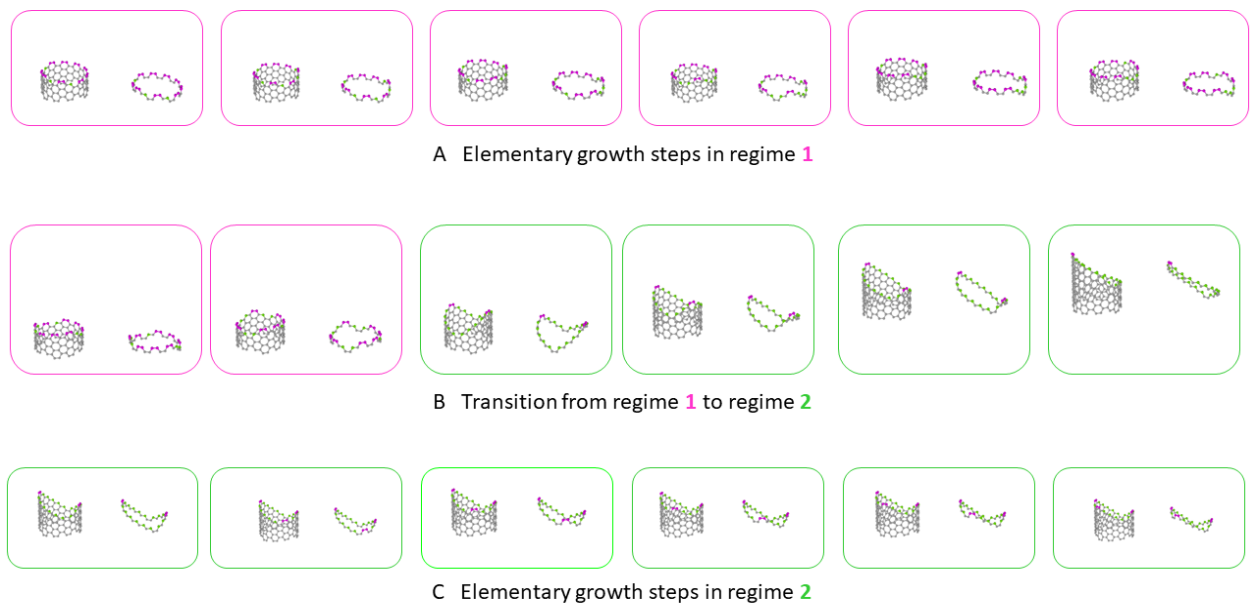

A   Elementary growth steps in regime **1**

B   Transition from regime **1** to regime **2**

C   Elementary growth steps in regime **2**

Snapshots of Kinetic Monte Carlo simulations showing elementary steps of the growth in regime 1(A) and 2 (C). A series of configurations selected during the transition from regime 1 to regime 2 is displayed in B. The transition takes place when (at least) two armchair pairs are set apart from each other, forming a saddle shape. One can easily understand that such a configuration induces a looser contact with a rigid catalyst (above the tube, not displayed in these images), or a strong distortion of the catalyst surface if it is more compliant.

**Figure S6**

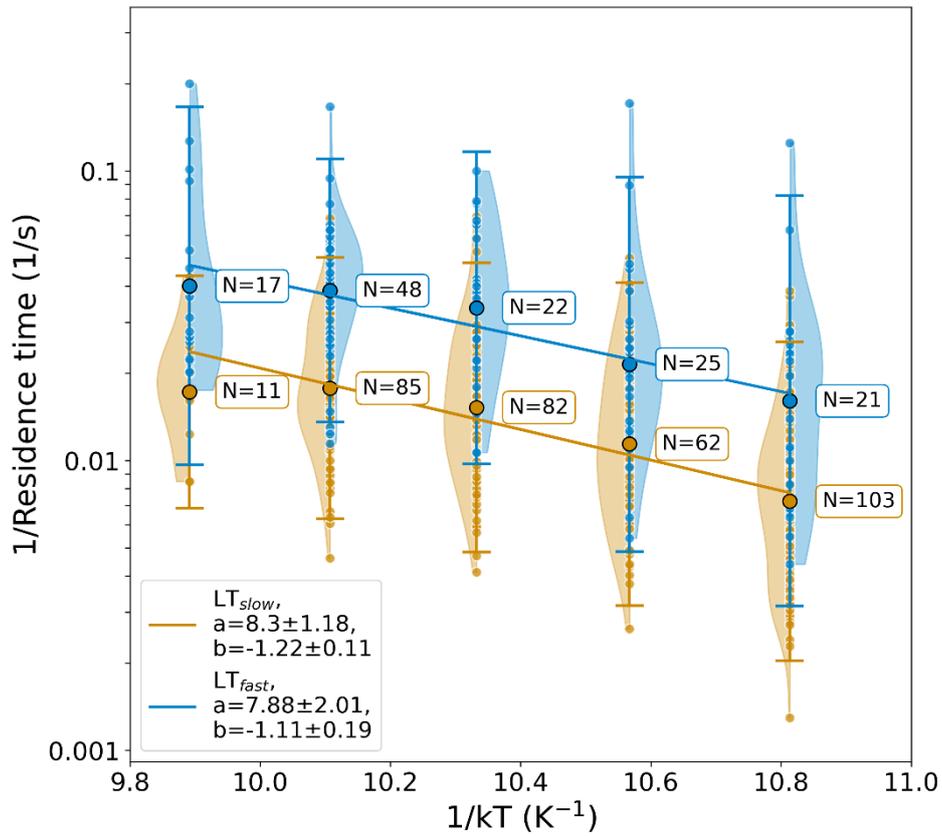

Jump frequencies of regimes 1 and 2, defined as the inverse of the experimental residence times in each regime. All segments followed by a transition, and only these, are included. Segments leading to an end of growth or any other event except another jump are therefore excluded. Different numbers of segments are therefore included in each regime.

The Arrhenius fit of these data yield values of $\Delta_1 = E_{max} - E_1 = 1.11 \pm 0.19$ eV and $\Delta_2 = E_{max} - E_2 = 1.22 \pm 0.11$ eV (defined in Fig. 3A), averaged over the experimental chiral distribution, which are used to set order of magnitude of the correction term introduced in the interfacial energy to create a bistability (see Eq. 7 , main text).

**Figure S7**

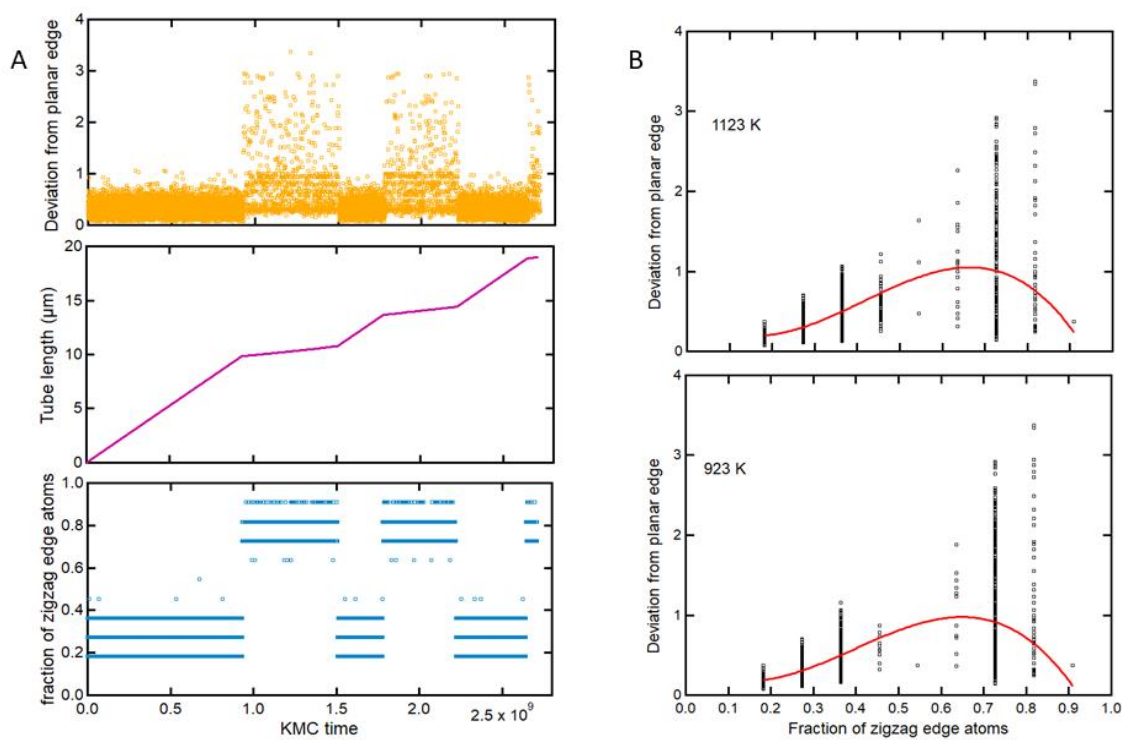

A) Example of transitions of a (13, 9) tube showing switches between regimes 1 and 2. Bottom panel: fraction of zigzag edge atoms characterizing the growth regime (< 0.5: regime 1; > 0.5: regime 2). Central panel: tube length in millimeters, as a function of KMC time. Top panel: characterization of the flatness of the tube / catalyst interface, using the mean squared distance to the plane that is closest to the edge atoms (in Å$^2$) as a parameter. In regime 1, the fraction of zigzag edge atoms is lower than 0.5. and the tube/catalyst interface is almost flat, with some roughness induced by the fluctuations of the edge structure during growth. In regime 2, this parameter is larger, with larger fluctuations.

B) Deviations from planar edge as a function of the fraction of zigzag edge atoms at 923 K and 1123 K. Out of the 2 10$^6$ KMC simulation steps, 25000 regularly spaced data points were stored and plotted here in black. Red curves are 3$^{rd}$ degree polynomial fits to these data. They are very similar at both temperatures. The central part which corresponds to the transition between the two regimes, shows a maximum deviation from flatness due to the saddle shape of the edge, related to the separation of the 2, 3 or 4 armchair pairs at the interface.

**Figure S8**

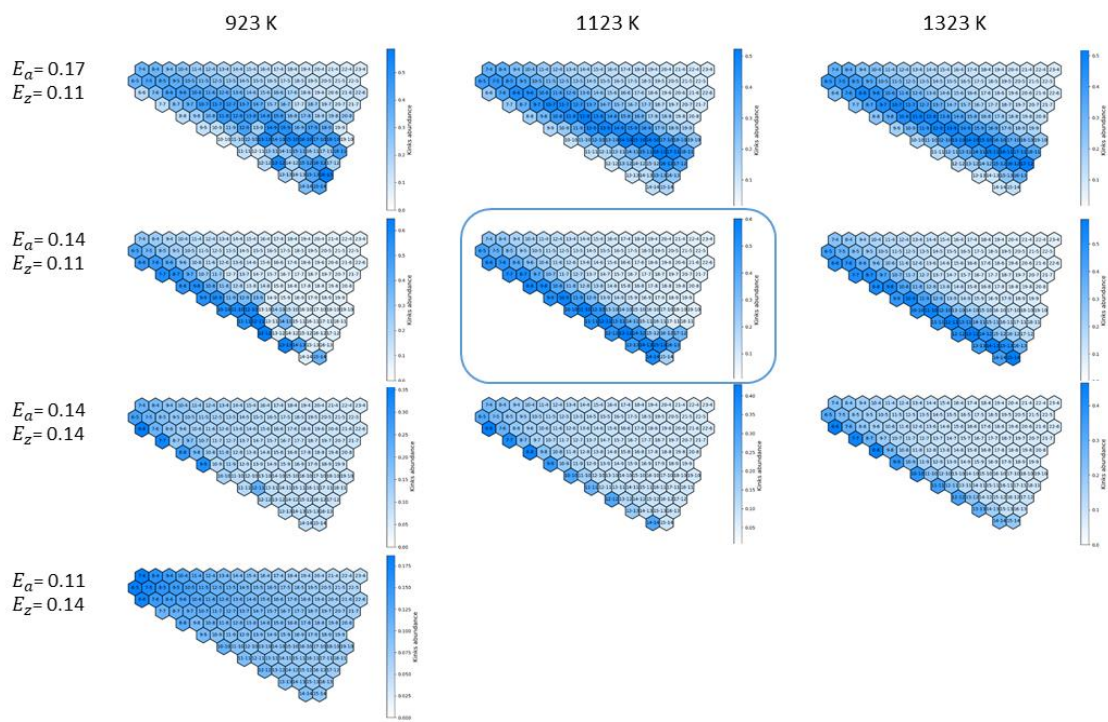

Maps displaying the number and amplitude of kinks for different parameters $E_A$, $E_Z$ (in eV), and T, using the standard deviation of the difference between the fractions of armchair and zigzag edge atoms ($N_Z - 2N_A$) as an indicator. The central panel, highlighted, matches the experimental data of growth rates for P=533 Pa and T=1123 K. Different trends can be observed. Increasing temperatures leads to a softening of the number and/or intensity of the kinks. For $E_A - E_Z \leq 0$, kinks tend to disappear, and all tubes grow in Regime 1. For usual catalysts (Fe, Co, Ni …), the reverse is generally observed ($E_A - E_Z > 0$), which means that broken line kinetics should be expected. If this difference is large enough, the critical domain of growth instabilities shifts towards smaller chiral angles. Tubes with chiral angles larger than the critical ones tend to grow in Regime 2, those with smaller chiral angles grow in Regime 1. Since the interface energies driving this simple modeling are not exactly known under real CVD conditions, these behaviors might appear as random and detrimental to a controlled growth.

**Figure S9**

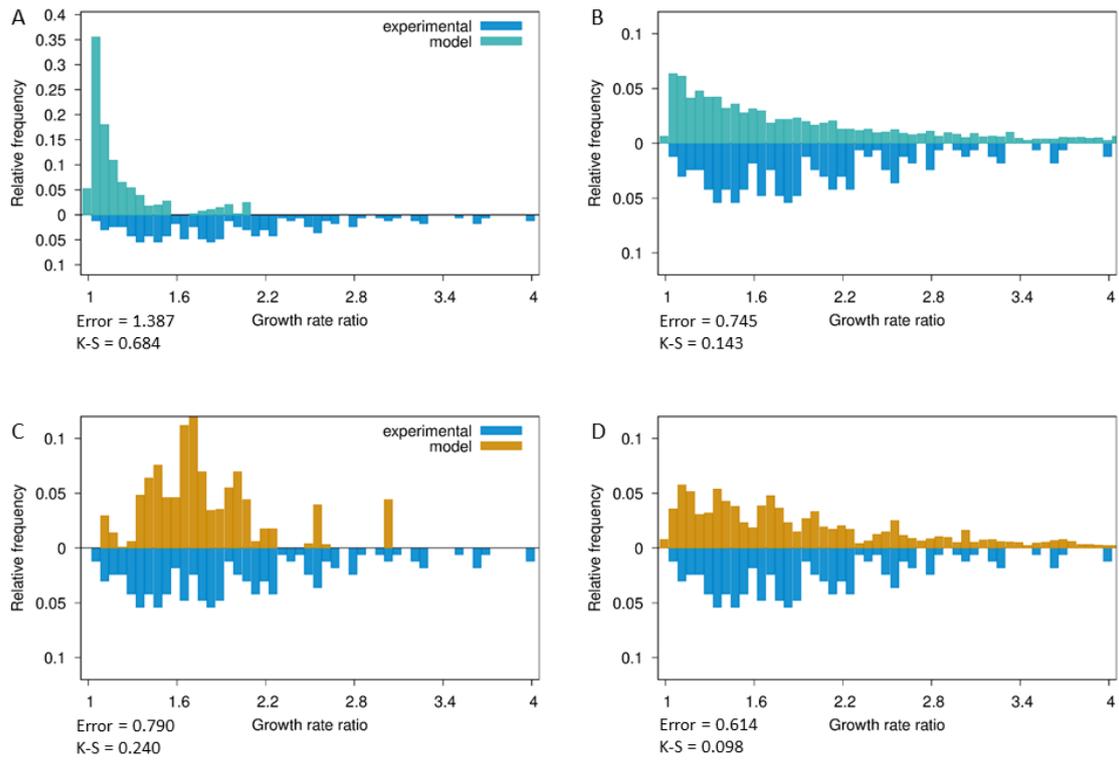

Comparison of the experimental distribution of growth rate ratios in the case of stochastic chirality switches with theoretical models. (A, B) Model with growth rate proportional to chiral angle. (C, D) Model assuming that chirality switches are accompanied by a change of kinetic regime. Two cases of chirality switches were considered: (A, C) all possible changes from (n, m) to its 6 surrounding chiralities, obtained by changing n or m by ± 1; (B, D) all possible chirality changes between CNTs of diameters within ± 2.5 Å from the reference one. The error and K-S values correspond to the total absolute difference and the Kolmogorov–Smirnov test between the model and the experimental data.

| P (Pa) | T (K) | Nb of events | Mean G1 | Mean G2 | Mean G1 / Mean G2 | Tr1 | Tr2 | Tr2/ Tr1 |
|---|---|---|---|---|---|---|---|---|
| 59 | 1123 | 90 | 0.239 | 0.150 | 1.59 | 91 | 108 | 1.19 |
| 178 | 1123 | 35 | 0.341 | 0.195 | 1.75 | 65 | 99 | 1.53 |
| 533 | 1073 | 124 | 0.162 | 0.093 | 1.74 | 92 | 153 | 1.67 |
| 533 | 1098 | 87 | 0.179 | 0.110 | 1.64 | 64 | 92 | 1.44 |
| 533 | 1123 | 104 | 0.301 | 0.179 | 1.68 | 48 | 69 | 1.44 |
| 533 | 1148 | 133 | 0.366 | 0.187 | 1.96 | 36 | 54 | 1.49 |
| 533 | 1173 | 28 | 0.436 | 0.274 | 1.59 | 38 | 46 | 1.18 |
| 1600 | 1123 | 466 | 0.808 | 0.407 | 1.99 | 18 | 28 | 1.56 |

**Table S1.** Summary of the 1067 growth rate changes analyzed. The pressure, temperature and numbers of rate change events for each P and T conditions are given. The next columns display average values of $G_1$, $G_2$ in $\mu m/s$, the ratio $R$, the residence times, $t_1^r$ and $t_2^r$ in seconds, their ratio, respectively. The data at P = 533 Pa, detailed in Data S1, were used to adjust the parameters of the Kinetic Monte Carlo simulations.

| Pressure (Pa) | Temperature (K) | Distribution of G1/G2 | | | Linear fit of G1 vs G2 | |
|---|---|---|---|---|---|---|
| | | Mean | Standard error of the mean (SEM) | Median absolute deviation | Slope | Standard error of the slope |
| 59 | 1123 | 1,70 | 0,08 | 0,26 | 1,38 | 0,05 |
| 178 | 1123 | 1,61 | 0,09 | 0,15 | 1,46 | 0,06 |
| 533 | 1073 | 1,87 | 0,04 | 0,28 | 1,64 | 0,03 |
| 533 | 1098 | 1,93 | 0,08 | 0,34 | 1,64 | 0,04 |
| 533 | 1123 | 1,84 | 0,10 | 0,24 | 1,58 | 0,03 |
| 533 | 1148 | 2,04 | 0,07 | 0,32 | 1,79 | 0,05 |
| 533 | 1173 | 1,65 | 0,07 | 0,16 | 1,49 | 0,05 |
| 1600 | 1123 | 2,18 | 0,05 | 0,40 | 1,81 | 0,03 |
| Average | | 1,85 | | | 1,60 | |
| | | | | | | |
| All pressures (T = 1123 K) | | 2,03 | 0,04 | 0,35 | 1,71 | 0,02 |
| All temperatures (P = 533 Pa) | | 1,91 | 0,04 | 0,29 | 1,65 | 0,02 |
| All data | | 2,00 | 0,03 | 0,33 | 1,71 | 0,02 |

**Table S2.** Assessment of the proportionality factor between experimental G1 and G2 values by the two methods (mean ratio, linear fit) for each dataset.